# Passively stable 0.7-octave microcombs in thin-film lithium niobate microresonators

**Zexing Zhao [1], Chenyu Wang [1], Jingyuan Qiu [2], Zhilin Ye [2], Zhijun Yin [2], Kunpeng Jia [1, *], Xiaohui Tian [1, **], Zhenda Xie [1, ***], and Shi-Ning Zhu [1]**

[1]National Laboratory of Solid State Microstructures, School of Electronic Science and Engineering, School of Physics, College of Engineering and Applied Sciences, and Collaborative Innovation Center of Advanced Microstructures, Nanjing University, Nanjing 210093, China

[2]NanZhi Institute of Advanced Optoelectronic Integration Technology Co., Ltd, Nanjing 210093, China

*Corresponding author: jiakunpeng@nju.edu.cn; **Corresponding author: tianxiaohui@nju.edu.cn; ***Corresponding author: xiezhenda@nju.edu.cn

## Abstract

Optical frequency comb based on microresonator (microcomb) is an integrated coherent light source and has the potential to promise a high-precision frequency standard, and self-reference and long-term stable microcomb is the key to this realization. Here, we demonstrated a 0.7-octave spectrum Kerr comb via dispersion engineering in a thin film lithium niobate microresonator, and the single soliton state can be accessed passively with long-term stability over 3 hours. With such a robust broadband coherent comb source using thin film lithium niobate, fully stabilized microcomb can be expected for massive practical applications.

## Introduction

As a highly coherent light source, the optical frequency comb generated on the microresonator (so-called microcomb) has many applications thanks to its high integrity, low power consumption, and low phase noise [1, 2]. Especially, octave spanning microcombs via dispersion engineering can realize a chip-scale 2f-3f or f-2f [3, 4] self-referencing scheme. This makes it a potential new frequency standard because of its high precision [5, 6]. In practice, the stability of the soliton comb source is the basis of subsequent signal processing. However, achieving a long-term stable soliton comb can be challenging due to thermal effects or center frequency jitter induced by the pump. This often requires a complex feedback system, which hinders the minimization of the device [7-10].

In recent years, thin-film lithium niobate (LN) photonic devices have been greatly developed and studied because of the broadband low-loss optical transparent window, excellent quadratic nonlinear effect, and high electro-optic efficiency [11-18]. In particular, the optical intensity-dependent photorefractive effect of LN [19-20] is just opposite to the thermo-optic nonlinear effect, so it enables a new mechanism for effectively generating soliton optical frequency combs based on LN microresonator with self-starting property [21-22]. Furthermore, such feature is potentially beneficial for long-term passive stability of the soliton comb, which is of significance for practical applications.

We report soliton microcombs generated in z-cut LN-on-insulator (LNOI) thin film micro-ring resonator (MRR), and different states of soliton micro-comb generation are observed. Moreover, the balance of photorefractive and thermo-optical effects of LN facilitates the generation of passively stable single soliton for more than 3 hours without feedback control. Finally, a wideband Kerr comb is generated in MRR with a smaller size based on dispersion engineering. We demonstrated the potential of high-performance LN microresonators as integrated, highly robust, and broadband coherent laser sources on a chip.

## Designs and fabrications

The High-Q LNOI resonators are fabricated by a standard E-beam lithography process. We initially use the electron beam to pattern the device on photoresist, and then the waveguide is fabricated by an Ar-ion milling process. Figure 1. (a)-(b) shows an optical image of the MRR device and the cross-section of the waveguide. The LN thin film (0.62 µm thickness) is partially etched down by 0.32 µm and the sidewall angle (θ) is estimated to be about 70°. The entire device is air-clad. We prepare the LN MRRs with different radii of 100 µm and 60 µm, corresponding to the free spectral ranges (FSRs) of 200 GHz and 333 GHz, respectively. We obtain anomalous dispersion in a wide spectral band by optimizing the geometry of the waveguide [23]. We used the single-layer inverse taper as the end-coupling structure with an estimated coupling efficiency of 13%. We simulate the $D_{int}$ curve (Fig. 1. (c), left, blue) of the MRR with a radius of 100 μm, the zero $D_{int}$ matching points are around 135 THz and 270 THz, identified as the position of DW. The inset is a zoom-in view of the black dashed box, and the experimentally extracted data are marked with orange circles.

## Experiments and Discussions

The schematic of the experimental setup is shown in Fig. 2. (a). The continuous wave laser is amplified by an erbium-doped fiber amplifier and coupled into the cavity through lens fiber as the pump light. We first measure the MRR transmission spectrum in the telecom band. The pump polarization is adjusted to match the TE cavity mode by the polarization controller (PC). The output optical signal is detected by a photodetector and followed by an oscilloscope to record the cavity mode transmission. The result is shown in Fig. 2 (b). Fig. 2. (c) is the zoom-in view of the selecting cavity mode and the Lorentzian fitting (red curve) indicates an extracted $Q_L$ over 1 million. As the power increases, we observe the appearance of soliton steps when the scanning pump light frequency decreases as shown in Fig. 2. (d). The output microcomb spectrum is measured by an optical spectrum analyzer (Fig. 2. (a)). Furthermore, we use the grating filter to suppress the pump mode. The extracted comb power is characterized by an electrical spectrum analyzer and optical power meter.

Different soliton comb states can be achieved by manually changing the pump frequency with the spectra shown in Figs. 3(a)-(h). Figure 3. (d) shows a single soliton comb with a repetition rate of 200 GHz, and the spectral envelope is fitted by the sech$^2$ function (red). However, we notice that the measured comb spectra feature no dispersive wave, which is perhaps caused by insufficient on-chip power (~260 mW). In addition, we also observe the two-soliton-state combs (Figs. 3. (e), (f)). Moreover, the low-frequency RF signal of generated combs in each state is measured to confirm the low-noise soliton state [22].

Figure 3. (i)-(j) shows the change of comb power during the generation of a single soliton comb. We first tune the frequency of the CW laser outside the cavity mode (Fig. 3. (i) I region) and increase it continuously until it enters the cavity and excites the four-wave mixing process (Fig. 3. (i) II region). We then use a bi-directional scanning scheme [21] to control the laser frequency until stable single solitons are generated (Fig. 3. (i)-(j) III region). The long-term stability of microcomb is a key parameter for practical applications. Figure 3. (j) shows the optical power of a single soliton state as a function of time. The single soliton state can be preserved for over 3 hours, showing excellent passive stability. In the process of microcomb generation in the LN microresonator, the thermal effect detunes the cavity mode toward red-region while the photorefractive effect of LN detunes it toward the blue-region [19, 20]. Therefore, the photorefractive effect can compensate for the thermal effect of the microresonator and facilitate a stable soliton microcomb state [20, 21].

The comb span is limited by the on-chip power in the 200 GHz MRRs, and we use 333 GHz MRR for the broadband comb generation. With similar on-chip power the circulating pump power is higher due to the higher finesse at the same Q. Figure 4. (a) is the simulated $D_{int}$ curve in this case of geometric parameters, the designed dispersive waves are located at 1315 nm and 2120. At the pump power of ~220 mW, we observe the broadband microcomb generation (Fig. 4. (b)). The dispersive waves generated by the Cherenkov radiation are located at 1250 nm and 2010 nm, respectively, the range of the comb line is from 1230 nm to 2100 nm (~0.7-octave). We find that the measured positions of DW deviated from the simulation and occurred blue-shift, while it could also be found in Figure 3 (c)-(g). Because LN is highly Raman active, the existence of Raman scatters will eventually produce a Raman-Kerr optical frequency comb in soliton forming a dynamic process, which causes additional energy transfer [20] and changes the condition for zero Dint point of DW. Furthermore, the broadband Kerr optical frequency comb via dispersion engineering can be further exploited such as 2f-3f reference technology research. For the 200 GHz MRRs, it is expected that the double-layer taper structure could provide sufficient on-chip power to realize the generation of an octave soliton comb, which can be used for an integrated all-optical self-reference locking scheme.

## Conclusions

In conclusion, we have demonstrated soliton microcombs generated on the lithium niobate platform based on the photorefractive effect. We fabricate high Q z-cut LN MRR by using a standard E-beam lithography process. In the experiment, a simple single-layer taper is used for end coupling with a measured coupling efficiency of 13%. We test the stability of the single soliton state in the pure free-running setup and the results show that the single soliton state can be passively stabilized for over 3 hours. Furthermore, the 0.7-octave spanning Kerr comb is observed with high stability, which can be expected to be used in 2f-3f self-referencing technology. In addition, the rich optical properties of LNOI make it possible for fully integrated self-referenced optical frequency comb realization. For example, the excellent electro-optic coefficient of LN can be used to generate microcomb with high-speed modulation capability, providing extra freedoms for dynamic control and active feedback for integrated microcomb sources. Such properties can be further applied to large-scale optical communication, microwave photonic technology, and other fields.

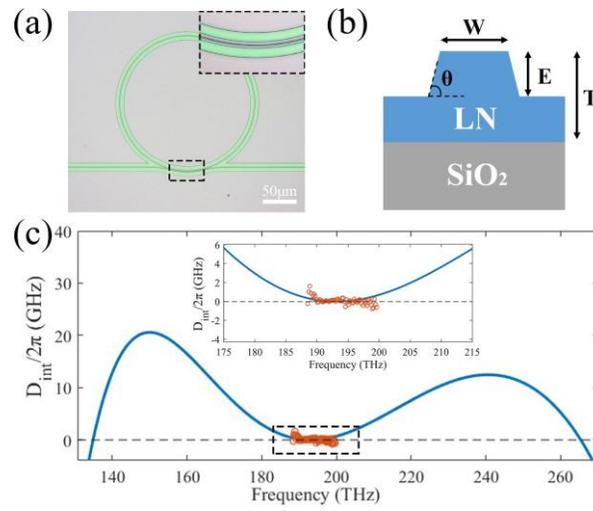

**Fig. 1.** (a) The optical image of the LNOI micro-ring resonator. Inset: zoom-in of the coupling region. (b) Cross-sectional schematic of the LN waveguide, with W (width) of 1.6 μm, E (etch depth) of 0.32 μm, and T (thickness) of 0.62 μm. (c) The simulated LN waveguide integrated dispersion $D_{int}/(2\pi)$ (blue). Inset: locally zoom-in within the black dashed box with the experimentally extracted data points (orange circles).

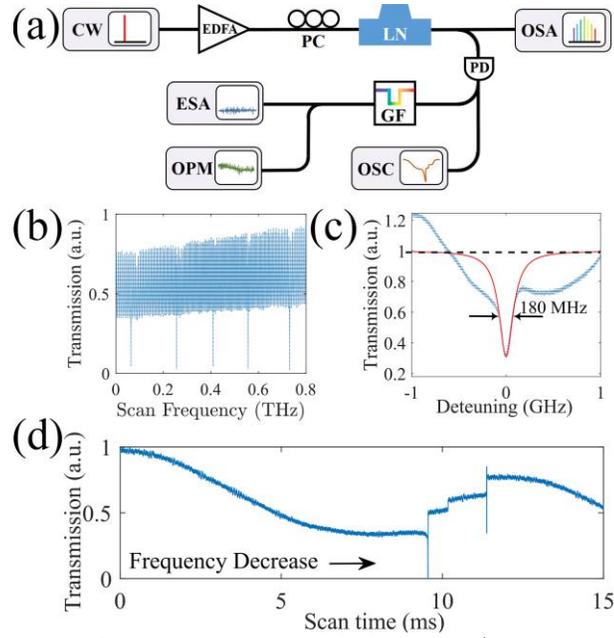

Fig. 2. (a) The schematic of experimental testing setups, including CW (continuous wave), EDFA (erbium-doped fiber amplifier), PC (polarization controller), PD (photodetector), OSC (oscilloscope), OSA (optical spectrum analyzer), ESA (electrical spectrum analyzer), GF (grating filter), and OPM (optical power meter). (b) The MRR transmission spectrum. (c) The laser-cavity detuning of the pump resonance mode (blue dot) with fitting Lorentz curve (red). (d) The resonator transmission at a high-power pump when the laser frequency decreased.

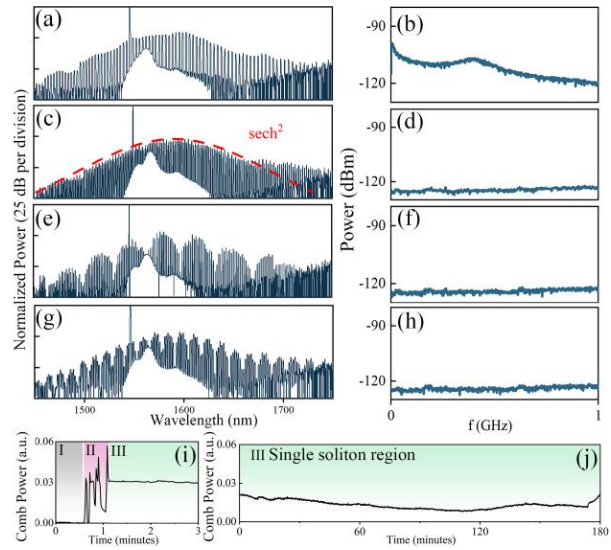

**Fig. 3.** The characterization of frequency combs generated on LN microresonator (radius 100 μm). The frequency comb spectra in different states are characterized, including (a) chaos comb, (c) single-soliton comb, and (e), (g) two-soliton combs with different relative angles between solitons. The mode-locked single-soliton comb was featured with a $sech^2$ fitting curve ((c), red dot line). Furthermore, the intensity noise of all frequency comb was characterized and shown in the right panel corresponding to each graph. The stable generation of single-soliton for long-term is realized by photorefractive and thermo-optical effects ((i)-(j)).

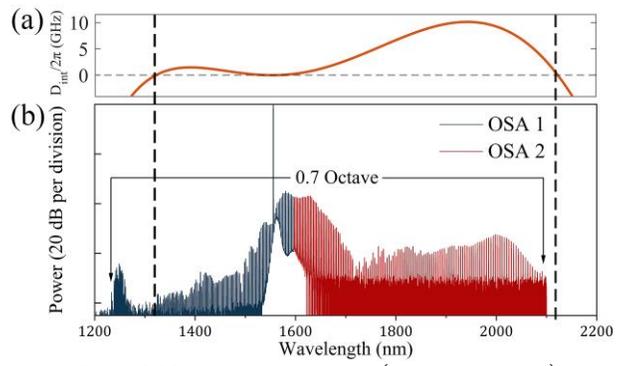

**Fig. 4.** The spectral characteristics of the LN microresonator (radius 60 µm) as well as simulation, the black dashed line marks the position of the dispersive wave on experimental results. (a) The simulated $D_{int}$ curves. (b) The normalized optical spectrum of the wideband Kerr comb.